\documentclass{ws-procs9x6}

\newcommand{\De}{\Delta}
\newcommand{\Sg}{\Sigma}
\newcommand{\Sgs}{\Sigma^*}

\newcommand{\be}{\begin{equation}}
\newcommand{\ee}{\end{equation}}
\newcommand{\Ld}{\Lambda(1520)}
\begin{document}

\title{Dynamical generation of $J^P=\tfrac{3}{2}^-$ resonances and the
$\Lambda(1520)$ resonance}

\author{Sourav~Sarkar, L.~Roca, E.~Oset, V.~K.~Magas and M.~J.~Vicente~Vacas}

\address{Departamento de F\'{\i}sica Te\'orica and IFIC,
Centro Mixto Universidad de Valencia-CSIC, \\ 
Investigaci\'on de Paterna, Aptdo. 22085, 46071 Valencia, Spain\\ 
}

\maketitle

\abstracts {
The lowest order chiral Lagrangian is used to study $s$-wave interactions of the
baryon decuplet with the octet of pseudoscalar mesons. The coupled channel Bethe
Salpeter equation is used to obtain dynamically generated
$\tfrac{3}{2}^-$ resonances in the partial wave amplitudes 
which provide a reasonable description to a number
of 3 and 4-star resonances like $\Delta(1700)$, $\Lambda(1520)$, $\Sigma(1670)$, 
$\Sigma(1940)$, $\Xi(1820)$, $\Omega(2250)$  etc. The 
phenomenological introduction of $d$-wave channels in the coupled channel 
scheme along with the existing $s$-wave channels
is shown to provide a much improved description of the $\Lambda(1520)$
resonance. The prediction of the absolute strength of the cross section in
the reactions $K^-p\to\Lambda\pi^0\pi^0$ and $K^-p\to\Lambda\pi^+\pi^-$
provided by this scheme shows a good agreement with existing  data. 
}

\section{Introduction}
The introduction of unitary techniques in a chiral dynamical treatment of the
meson baryon interaction has been very successful. It has lead to good
reproduction of meson baryon data with a minimum amount of free parameters, and
has led to the dynamical generation of many low lying resonances which qualify
as quasibound meson baryon states~\cite{kaiser,ramos}. In particular,
the application of these techniques to the $s$-wave 
scattering of the baryon octet and the pseudoscalar meson octet have led to the
successful description of many $J^P=\frac{1}{2}^-$ resonances like the
$N^*(1535)$, the $\Lambda(1405)$, the $\Lambda(1670)$ and the 
$\Sigma(1620)$~\cite{inoue,jido,bennhold,Garcia-Recio:2002td}. Naively one may expect that this scheme is not suitable for
studying $d$-wave resonances due to a large number of unknown parameters in
the corresponding chiral Lagrangian. However $d$-wave resonances could be 
studied in $s$-wave interactions of the meson octet with the baryon
decuplet~\cite{lutz,Sarkar:2004jh,VicenteVacas:2005dn}, in
which case chiral dynamics is quite predictive. Applying unitary
techniques to the lowest order chiral Lagrangian we have been successful in 
generating a number of $\tfrac{3}{2}^-$ resonances. From the information of
the pole positions and couplings to the channels involved we could associate many of these
resonances to the $N^*(1520)$, $\De(1700)$, $\Lambda(1520)$, $\Sigma(1670)$,
$\Sigma(1940)$, $\Xi(1820)$ resonances tabulated by the Particle Data Group (PDG). 

We then consider the dynamics of one of the above resonances, namely
the $\Lambda(1520)$, which is capturing renewed
attention, particularly since it appears invariably in searches
of pentaquarks in photonuclear reactions like $\gamma p\to K^+K^-p$ and $\gamma
d\to\ K^+K^-np$. In the simple picture~\cite{Sarkar:2004jh}
 mentioned above, this resonance couples to the $\pi \Sgs(1385)$ and 
$K \Xi^*(1530)$ channels, particularly to
the former one. 
With the $\pi^+ \Sg^{*-}$, $\pi^- \Sg^{*+}$, $\pi^0 \Sg^{*0}$ masses
7 MeV above, 2 MeV above and 1 MeV below the nominal $\Lambda(1520)$ mass and the
strong coupling of the resonance to $\pi \Sgs$, 
this state could qualify as a loosely 
bound $\pi \Sgs$ state. 
  However, the lack of other relevant channels which couple
to the quantum numbers of the resonance makes the treatment 
of~\cite{Sarkar:2004jh} only semiquantitative. In
particular, the $\Lambda(1520)$ appears 
at  higher energy than the nominal one and with a  large width of
about 130 MeV, nearly ten times larger than the physical width.
  This large width is a necessary consequence of the large coupling to the 
$\pi \Sgs$ channel and the fact that the pole appears at energies above
the $\pi \Sgs$ threshold. On the other hand, if we modify the
subtraction constants of the meson baryon loop function to bring the pole below
the $\pi \Sgs$ threshold, then the pole appears without imaginary part.
Since the width of the $\Lambda(1520)$ resonance comes basically from the
decay into
the $\bar{K} N$ and $\pi \Sigma(1193)$, the introduction of these channels is 
mandatory to reproduce the shape of the $\Lambda(1520)$ resonance.

In~\cite{Sarkar:2005ap,luis_prep} we phenomenologically include the $\bar{K} N$ and $\pi \Sigma$ channels into
the set of coupled channels which build up the $\Lambda(1520)$.
 The novelty with
respect to the other channels already accounted for~\cite{Sarkar:2004jh},
 which couple in $s$-wave,
 is that the new channels couple in $d$-waves. The transition matrix elements 
 for the $d$-wave channels $\bar K N$ and $\pi\Sg$ to the $s$-wave channel
 $\pi\Sgs$ were parametrized
 in terms of unknown constants 
 which were fitted to experimental values for the real and imaginary parts of 
 the partial wave
 amplitudes for the reactions 
 $\bar K N\to\bar K N$ and $\bar K N\to\pi\Sigma$.
 The coupling of the $\Lambda(1520)$ to the $\pi \Sgs$ 
channel is then a prediction of this scheme and we use this to study
the reaction $K^- p \to \pi^0 \pi^0\Lambda$ and $K^- p \to \pi^+ \pi^-\Lambda$
which are closely related to the strength of 
this coupling. The prediction of the absolute strength of the cross sections in
 the above two reactions close to and above the $\Lambda(1520)$ energy
find a good agreement with  data.

\section{Dynamical generation of spin 3/2 baryon resonances}

The tree-level scattering amplitude involving the baryon decuplet and the
pseudoscalar octet is obtained from the dominant lowest order chiral Lagrangian 
which accounts for the Weinberg Tomozawa term.
The matrix containing these amplitudes, $V$ is used as the kernel of a coupled
channel Bethe Salpeter equation given by
\be
T=V+VGT
\label{eq:bethe}
\ee
to obtain the transition matrix fulfilling exact unitarity 
in coupled channels. The factor $G$ in the above equation
is the meson baryon loop function. 
We have looked in detail~\cite{Sarkar:2004jh,VicenteVacas:2005dn} at the $\frac{3}{2}^-$ 
resonances which are generated dynamically by this interaction, by searching for
poles of the transition matrix in the complex plane in different Riemann sheets.
\begin{figure}[t]
\centerline{
\includegraphics[width=1.0\textwidth]{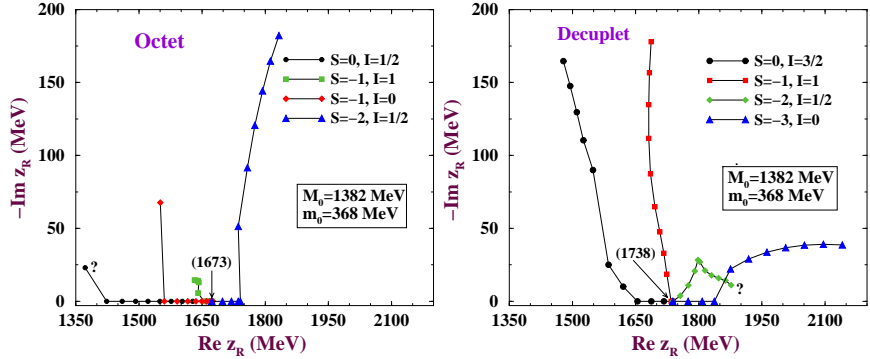}
}
\caption{Trajectories of the poles in the scattering amplitudes obtained by
increasing the value of the $SU(3)$ breaking parameter in steps of $0.1$
(symbols) from
zero, which is the $SU(3)$ symmetric situation, up to $1$ which corresponds to
the physical masses.}
\label{fig:trajfig}
\end{figure} 
We start from a $SU(3)$ symmetric situation where the masses of
the baryons are made equal and the same is done with the masses of the mesons. In this
case we found attraction in the octet, decuplet and the 27 representations,
while the interaction was repulsive in the 35 representation.  In the $SU(3)$
symmetric case all states of the $SU(3)$ multiplet are degenerate and the
resonances appear as bound states with no width. As we gradually break $SU(3)$
symmetry by changing the masses, the degeneracy is broken and the states with
different strangeness and isospin split apart generating pole trajectories in the
complex plane which
lead us to the  physical situation in the final point, as seen in
fig.~\ref{fig:trajfig}.  This systematic search
allows us to trace the poles to their $SU(3)$ symmetric origin, although there is
 a mixing of representations when the symmetry is broken. 

\begin{figure}[t]
\centerline{
\psfig{figure=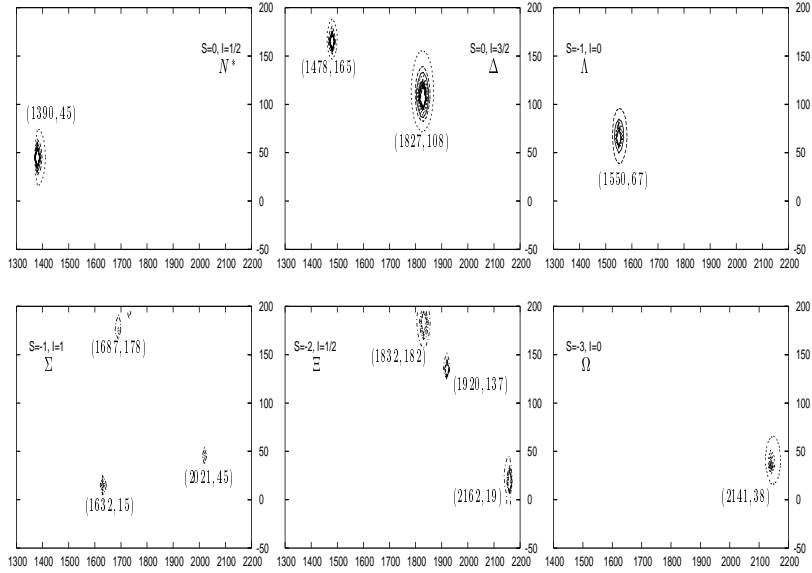,width=10.5cm,height=7.5cm,angle=-90}
}   
\caption{Contour plots of the scattering amplitudes showing poles
 in the unphysical Riemann sheets. The
$x$ and $y$ axes denote the real and imaginary parts of the CM energy.}
\label{fig:polesall}
\end{figure}
\begin{figure}[t]
\centerline{
\psfig{figure=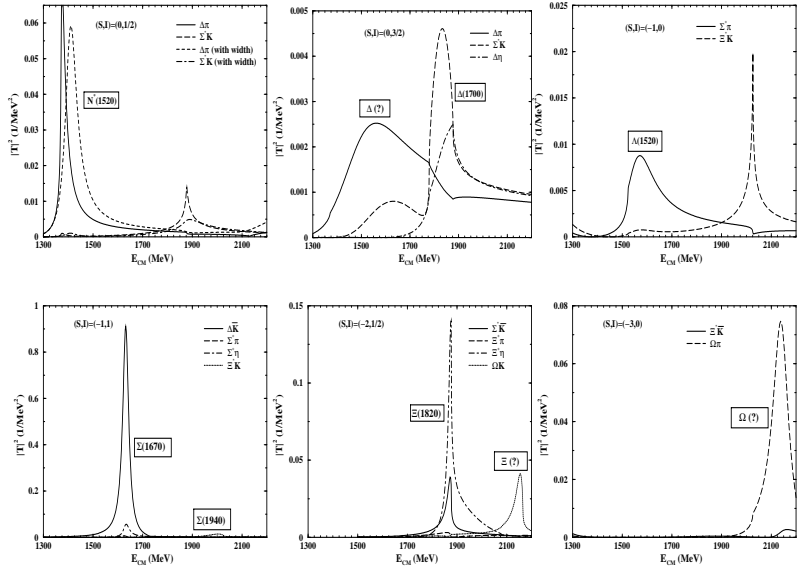,width=10.5cm,height=7.5cm}
}
\caption{Scattering amplitudes as a function of CM energy showing
 $3/2^-$ resonances generated dynamically in decuplet
baryon - pseudoscalar octet interactions.}
\label{fig:ampall}
\end{figure}

In fig.~\ref{fig:polesall} we show contour plots of the partial
wave amplitudes as a function of the complex CM energy. The projection on the
real (energy) axis is shown in fig.~\ref{fig:ampall}.
We have also evaluated the residues of the poles from where the couplings of the
resonances to the different coupled channels were found and this allowed us to
make predictions for partial decay widths into a decuplet baryon  and a meson.
There is very limited experimental information on these decay channels but,
even then, it represents an extra check of consistency of the results  which
allowed us to more easily identify the resonances found with some resonances
known, or state that the resonance should correspond to a new resonance not yet
reported by the PDG. In particular, in view of the information of
the pole positions and couplings to channels we could associate some of the
resonances found to the $N^*(1520)$, $\De(1700)$, $\Lambda(1520)$, $\Sigma(1670)$,
$\Sigma(1940)$ and $\Xi(1820)$.
\section{Improved description of the $\Lambda(1520)$ resonance}

In addition to the transition amplitudes involving the $s$-wave channels $\pi\Sgs$ and $K\Xi^*$ which were used to
dynamically generate the $\Ld$ as discussed in the previous section
we introduce phenomenologically tree level transition potentials 
involving the $d$-wave channels $\bar{K}N$ and $\pi\Sigma$.
As discussed in~\cite{Sarkar:2005ap}, we use for the vertices 
$\bar{K}N\to\bar{K}N$, $\bar{K}N\to\pi\Sigma$ and
$\pi\Sigma\to\pi\Sigma$ effective
transition potentials which are 
proportional to the incoming and outgoing momentum squared. 
 Denoting  $\pi\Sigma^*$, $K\Xi^*$,  $\bar K N$  and
$\pi\Sigma$ channels by $1$, $2$, $3$ and $4$ respectively, the
matrix containing the tree level amplitudes is written as:

\be
V=\left( 
\begin{array}{cccc}
C_{11}(k_1^0+k_1^0)\ & C_{12}(k_1^0+k_2^0) & \gamma_{13}\,q_3^2 &\gamma_{14}\,q^2_4 \\
C_{21}(k_2^0+k_1^0)\ & C_{22}(k_2^0+k_2^0) & 0 & 0 \\
\gamma_{13}\,q_3^2 & 0 & \gamma_{33}\, q^4_3 & \gamma_{34} \,q_3^2 \,q^2_4\\
\gamma_{14}\,q^2_4  & 0 & \gamma_{34} \,q_3^2 \,q_4^2 &  \gamma_{44}\, q^4_4
\end{array}
\right)~,
\ee
\noindent
where $q_i=\frac{1}{2\sqrt{s}}\sqrt{[s-(M_i+m_i)^2][s-(M_i-m_i)^2]}$,
$k_i^0=\frac{s-M_i^2+m_i^2}{2\sqrt{s}}$ 
and $M_i(m_i)$ is the baryon(meson) mass. 
The coefficients $C_{ij}$ are $C_{11}=\frac{-1}{f^2}$,
 $C_{21}=C_{12}=\frac{\sqrt{6}}{4f^2}$ and $C_{22}=\frac{-3}{4f^2}$,
where $f$ is $1.15f_\pi$, with $f_\pi$ ($=93$ MeV) the pion decay constant,
which is taken as an average between $f_\pi$ and
$f_K$. The elements $V_{11}$, $V_{12}$, $V_{21}$, $V_{22}$ come from the
lowest order chiral Lagrangian involving the decuplet of baryons and
the octet of pseudoscalar mesons~\cite{Sarkar:2004jh,lutz}.
We neglect the elements $V_{23}$ and $V_{24}$ which involve the
tree level interaction of the $K\Xi^*$ channel to the $d$-wave channels
because the $K\Xi^*$ threshold is far away from the $\Ld$ and its role
in the resonance structure is far smaller than that of the
$\pi\Sigma^*$.
It is also important to emphasize that the consideration of the width of the $\Sigma^*$ resonance
in the loop function $G$ is crucial in order to account properly for
the $\pi\Sigma^*$ channel since the threshold lies in the $\Lambda(1520)$
region. This is achieved through the convolution of the $\pi\Sigma^*$ loop
function with the spectral distribution considering the $\Sigma^*$
width.

In the model described so far we have as unknown parameters
$\gamma_{13}$, $\gamma_{14}$, $\gamma_{33}$, $\gamma_{34}$,
$\gamma_{44}$  in the $V$ matrix. Apart from these, there is also
the freedom in the value of the subtraction constants in the loop
functions. We will consider one subtraction constant for the
$s$-wave channels ($a_0$) and one for the $d$-wave ones ($a_2$).
Despite the apparent large number of free parameters in the $V$
matrix, it is worth emphasizing that the largest matrix elements
are $V_{11}$, $V_{12}$ and $V_{22}$ \cite{Sarkar:2004jh}
which come from a chiral  Lagrangian  without any
free parameters. Due to the $d$-wave behavior
the other ones are expected to
be smaller,  as we will see below.
In order to obtain these parameters we fit 
the partial wave amplitudes obtained by using the $V$ matrix given above
as the kernel in
the Bethe-Salpeter equation
to the
experimental results on the $\bar K N$ and $\pi\Sigma$ scattering
amplitudes in $d$-wave and $I=0$.
We use experimental data from~\cite{Gopal:1976gs,Alston-Garnjost:1977rs} where 
$\bar K N\to\bar K N$ and $\bar K N\to\pi\Sigma$ amplitudes are
provided from partial wave analysis.
These experimental amplitudes are related to the
amplitudes of Eq.~(\ref{eq:bethe}) through 
\(
\tilde{T}_{ij}(\sqrt{s})=-\sqrt{\frac{M_iq_i}{4\pi\sqrt{s}}}
\sqrt{\frac{M_jq_j}{4\pi\sqrt{s}}}\,T_{ij}(\sqrt{s})~,
\)
where $M$ and $q$ are the baryon mass and the on-shell 
C.M. momentum of the channel respectively. Note that in this normalization
the branching ratio is simply given by the strength of the imaginary part as
\(
B_i={\Gamma_i}/{\Gamma}=Im \tilde{T}_{ii}(\sqrt{s}=M_R).
\)

From the fit we obtain the subtraction constants $a_0=-1.8$ for the $s$-wave
channels and $a_2=-8.1$  for the $d$-wave channels. The unknown constants in the
$V$ matrix are given by $\gamma_{13}=0.98$ and $\gamma_{14}=1.1$ in units
of $10^{-7}$ MeV$^{-3}$ and $\gamma_{33}=-1.7$, $\gamma_{44}=-0.7$ and
$\gamma_{34}=-1.1$ in units of $10^{-12}$ MeV$^{-5}$.
We can see that the value obtained for the subtraction constant for
the $s$-wave channels is of  natural size ($\sim-2$) since it 
agrees with the result obtained
with the cutoff method using a cutoff of about $500$ MeV 
(at $\sqrt{s}\simeq 1520$ MeV).
 On the other hand, regarding the $d$-wave loops,
the large value obtained for $a_2$ can be understood comparing also
to the cutoff method. If one keeps the momentum dependence of the
 $d$-wave vertices inside the
loop integral (i.e., one does not use the on-shell approximation
mentioned above) and evaluates the integral with the cutoff method,
then also a cutoff of about  $500$ MeV  gives the
same result as the dimensional regularization with on-shell
factorization and $a_2\sim-8$.
In summary, the use of the dimensional regularization method
along with the on-shell factorization for both the $s$ and $d$-wave
loops, correspond to the result obtained with the cutoff method
without on-shell factorization using the same cutoff of about
$500$ MeV.

\begin{figure}[t]
\centerline{
\includegraphics[width=1.0\textwidth,angle=0]{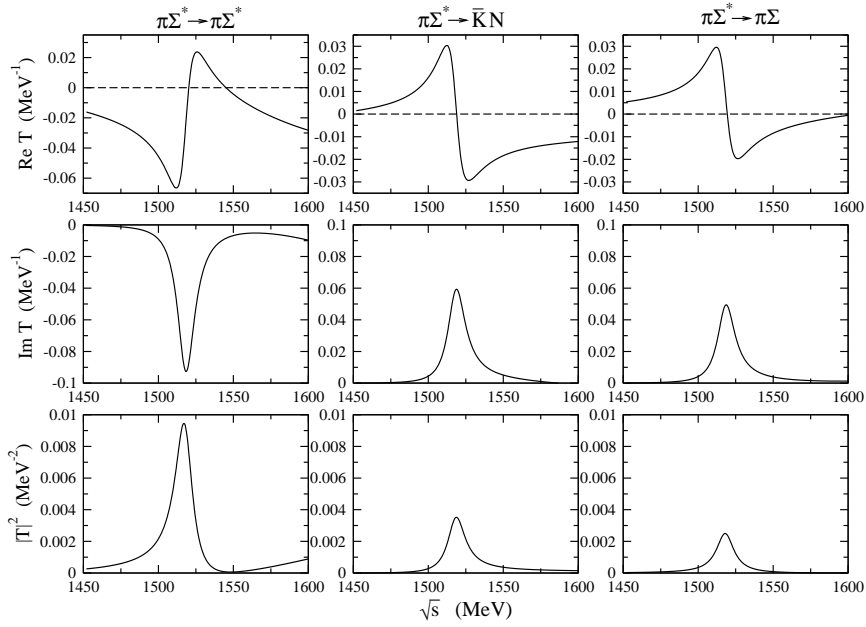}
}
\caption{Unitary amplitudes involving the $\pi\Sigma^*$ channel.
 From left to right:
 $\pi\Sigma^*\to\pi\Sigma^*$, $\pi\Sigma^*\to \bar K N$ and 
$\pi\Sigma^*\to \pi\Sigma$.}
\label{fig:Tij}
\end{figure}
We now show in fig.~\ref{fig:Tij}, the prediction for the
unitarized
amplitudes  for the different channels involving the $\pi\Sigma^*$.
From left to right the columns represent the
 $\pi\Sigma^*\to\pi\Sigma^*$, $\pi\Sigma^*\to \bar K N$ and 
$\pi\Sigma^*\to \pi\Sigma$ channels.
The rows denote from top to bottom the real part, imaginary part and modulus
squared of the amplitudes ($T_{ij}$) respectively.
We do not show the $K\Xi^*$ channel since it is less relevant as
an external state in physical processes. 

\begin{figure}[t]
\centerline{
\includegraphics[width=1.0\textwidth]{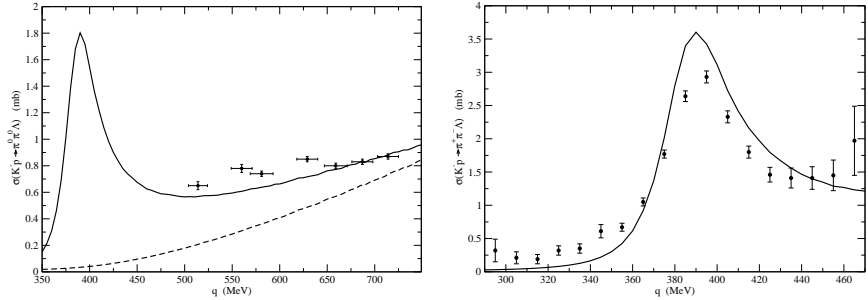}
}
\caption{Result for the $K^-p\to\pi^0\pi^0\Lambda$ (left) and 
$K^-p\to\pi^+\pi^-\Lambda$ (right) cross section.
}
\label{fig:nef_mast}
\end{figure}
From the imaginary part of the amplitudes it is straightforward to
obtain the couplings of the $\Ld$ to the different channels
in the following way. 
Close to the peak the amplitudes can be approximated 
by
\be
T_{ij}(\sqrt{s})=\frac{g_i g_j}{\sqrt{s}-M_{\Ld} + i \Gamma_{\Ld} /2}
\label{eq:Tcoup}
\ee 
from where we have
\be
g_i g_j=-\frac{\Gamma_{\Ld}}{2} \frac{|T_{ij}(M_{\Ld})|^2}
{Im[T_{ij}(M_{\Ld})]},
\ee
\noindent 
where $M_{\Ld}$ is the position of the peak in $|T_{ij}|^2$ and 
$\Gamma_{\Ld}=15.6$ MeV.
Up to a global sign of one of the couplings (we choose $g_1$ to be
positive), we obtain: $g_1=0.91$, $g_2=-0.29$, $g_3=-0.54$
 and $g_4=-0.45$ for the coupling of the $\Ld$ to the $\pi\Sgs$, $K\Xi^*$,
$\bar K N$ and $\pi\Sg$ channels respectively.
We can see from the values that the $\Ld$ resonance 
couples most strongly to the $\pi\Sgs$ channel.
The fact that we are able to predict the value of this coupling is a
non trivial consequence of the unitarization procedure that we
employ. With the value for $g_1$ obtained above,
we can evaluate the partial
decay width of the $\Lambda(1520)$ into $\pi\pi\Lambda$ 
assuming that this process is dominated by the $\pi\Sigma^*$ channel, and this
leads us to a branching ratio of 0.14. 
All these branching ratios (the branching ratios obtained for $\bar K N$
and $\pi\Sigma$ are $0.45$ and $0.43$ respectively)
 essentially
sum up to unity considering the uncertainties in the calculations.

The prediction of the amplitudes involving $\pi\Sigma^*$ channels can
be checked in particular reactions where this channel could play an
important role. We evaluate the cross section for $K^-p\to\pi\pi\Lambda$ in the
lines of~\cite{Sarkar:2005ap} but using the new coupled channel
formalism. The mechanisms and the 
expressions
 for the amplitudes and
the cross sections can be found in~\cite{Sarkar:2005ap} where,
apart from the coupled channel unitarized amplitude, other
mechanisms of relevance above the $\Ld$ peak were also included.
In fig.~\ref{fig:nef_mast} 
 we show our results for 
$K^-p\to\pi^0\pi^0\Lambda$ and $K^-p\to\pi^+\pi^-\Lambda$
 cross section on the left and right panels respectively along with experimental data
from refs.~\cite{Prakhov:2004ri} and ~\cite{Mast:1973gb} respectively.
The dashed line in the left figure  represents the contribution from mechanisms 
 other than the unitarized
coupled channels, and the solid line gives the coherent sum of all
the processes.
These cross sections depend essentially on the $T_{\bar K
N\to\pi\Sgs}$ amplitude which is obtained from our coupled channel
analysis. It is  a non-trivial prediction of the theory  since this
amplitude has not been included in the fit. 
\begin{figure}[t]
\centerline{
\includegraphics[width=0.6\textwidth]{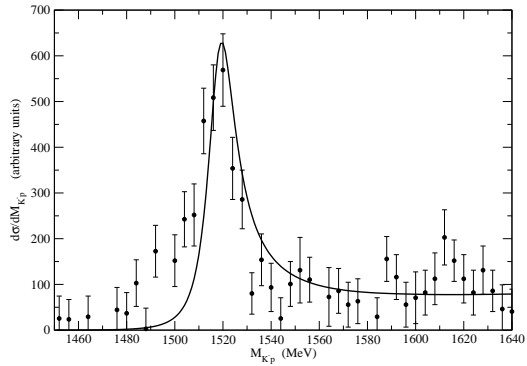}
}
\caption{$K^-p$ invariant mass distribution for the
 $\gamma p\to K^+ K^-p$ reaction with photons in the range
 $E_\gamma=2.8-4.8\textrm{ GeV}$.
Experimental data from ~\protect\cite{Barber:1980zv}.}
\label{fig:dares}
\end{figure}

We have also evaluated the $K^-p$ invariant mass distribution for the
$\gamma p\to K^+ K^-p$ reaction. In~\cite{Roca:2004wt} the basic
phenomenological model is explained but there only the $\pi\Sigma^*$
and $K\Xi^*$ channels were considered. The result in the present model
is shown in Fig.~\ref{fig:dares}. The normalization is arbitrary in the experimental data as well
as in our calculation. For the purpose of the present work the
shape of the distribution is the most important part and we can
see that the agreement is quite fair.

\section{Conclusions}

We have done a coupled channel analysis of the $\Ld$ resonance
using the  $\pi\Sigma^*$, $K\Xi^*$, $\bar{K}N$ and
$\pi\Sigma$ channels. We have used the Bethe-Salpeter equation
to implement unitarity  in the evaluation of the different 
amplitudes. The main novelty from previous coupled channel
approaches to this resonance is the inclusion of new matrix
elements  in the kernel of the Bethe-Salpeter equation and the consideration
of the $\Sigma^*$ width in the $\pi\Sigma^*$  loop function.
The unknown parameters in the $V$ matrix, as well as the
subtraction constants of the loop functions, have been obtained
by a fit to $\bar K N\to\bar K N$ and $\bar K N\to \pi\Sigma$
partial wave amplitudes. As a consequence of the unitarity of
the scheme used, we can predict the amplitudes and couplings of
the $\Ld$ for all the different channels.  The largest coupling
is obtained for the $\pi\Sgs$ channel.

We have then tested the amplitudes obtained in several specific
reactions and compared with experimental data at energies 
close to and slightly above the $\Ld$ region. These include the
$K^-p\to\Lambda\pi^0\pi^0$, $K^-p\to\Lambda\pi^+\pi^-$ and  
$\gamma p\to K^+K^-p$ 
  reactions.
We have obtained a reasonable agreement with the 
experimental results
that allows us to be confident in the procedure followed
 to describe the
nature of the $\Ld$ resonance.
\section*{Acknowledgments}
\small{This work is partly supported by DGICYT contract number BFM2003-00856,
and the E.U. EURIDICE network contract no. HPRN-CT-2002-00311.
This research is part of the EU Integrated Infrastructure Initiative
Hadron Physics Project under contract number RII3-CT-2004-506078.}

\end{document}